\documentclass[12pt]{article}
\usepackage{latexsym}
\usepackage{graphicx}

\begin{document}


\renewcommand{\thefootnote}{\fnsymbol{footnote}}

\title{\Large\bf Naive Drell-Yan and Its Successor\thanks{Talk given at the Drell Fest, July 31, 1998, SLAC on the occasion
 of Prof. Sid Drell's retirement.}}

\author{T-M. Yan \\
        Floyd R. Newman Laboratory of Nuclear Studies\\
        Cornell University\\
        Ithaca, NY  14853}
\maketitle


\abstract{We review the development in the field of lepton pair production
 since proposing parton-antiparton annihilation as the mechanism of massive
 lepton pair production.  The basic physical picture of the Drell-Yan model
 has survived the test of QCD, and the predictions from the QCD improved
 version have been confirmed by the numerous experiments  performed in the last
 three decades.  The model has provided an active theoretical arena for
 studying infrared and collinear divergences in QCD.  It is now so well
 understood theoretically that it has become a powerful tool for new phyiscs
 information such as precision measurements of the W mass and lepton and
 quark sizes.}

{\vspace*{-5.5 in}
\rightline{\textsf{CLNS 98/1580}}}
\vfill

\pagebreak

\indent{I am very happy to be here to participate in the festivities to celebrate Sid Drell's accomplishments.}

\indent{I met Sid for the first time exactly 30 years ago in July 1968 when I 
came to SLAC as a postdoc.  I actually met him and BJ in spirit a few months earlier.  I did my PhD with Julian Schwinger. And there was this story that Julian did not want his students to use Feynman diagrams.  I never found out if the story was true, since I didn't even know what a Feynamn diagram was!  I was quite alarmed when I noticed that everyone else was using Feynman diagrams in their calculation.  So after finishing my thesis in the Spring of 1968, I started to study carefully the old bible of quantum field theory, ``Bjorken and Drell.''  I was very glad I did that before I came to SLAC.  In the summer of 1968, the excitement at SLAC was unbelievable.  The data from SLAC-MIT experiments were coming out, BJ just proposed his scaling and Feynman invented the parton model right here at SLAC.  So I had to start doing research immediately and intensively.  Thus it was the beginning of a happy association with SLAC and a dear friendship with Sid and Harriet and the Drell family.}
 
\indent{In coming to this meeting, many fond memories and stories come alive again in my mind.  The first thing that suprised me most was Sid's phone calls to me in the evenings and on the weekends to discuss physics.  To understand my surprise, I have to provide you with a little background.  Julian Schwinger was a great phisicist and a wonderful teacher, and I loved him dearly.  But I would never expect any phone calls from him on weekend or otherwise.  Sid and I had many productive weekend discussions on physics.  Those were exciting times in my research career.}

\indent{SLAC was a wonderful place to work, but it was also a very nice place to relax.  I liked to put up my legs on the desk to goof off.  Sid thought I was in deep thought on something profound in physics.  Obviously, I did not want to countradict him.  But then, once in a while, I was actually sitting up in my office.  On those rare occasions, Sid would say to me,``Tung-Mow, you are not working!''  Those were the happy good old days.

\indent{Sid really cares about people, especially the young people, like students and postdocs.  He would do everything possible to advance our careers.  I know he had done many things to help me.  I am also certain that he did many more things for me that I did not know about.  Sid did everything, large and small, to make lives a little easier for us so we could work a little harder on physics, and at the same time to enjoy it!  One example could be found in his old office. That office has two doors.  Do you ever wonder why this is so?  The reason is Sid was concerned if there was only one door students and postdocs who wanted to see him would have to go through his secretary Bobby and that could be a little intimidating (Bobby was nice and kind, not really intimedating at all).  He wanted to have a second door so we could sneak into his office without going through the secretary.  This was all fine, except when he received confidential phone calls.  Then he had to close the two doors simultaneously!}

\indent{After 30 years, Sid's children have grown up and so have mine.  One of Sid's daughter, Persis, is now a physics professor at Cornell.  By the way Persis is not here today because her father forbade her to come.  According to Persis, this is the first time she obeyed her father.   A few years ago, my younger son Tony was a physics major at Cornell.  I thought to myself: Wouldn't it be nice and amusing to have another Drell teach another Yan about Drell-Yan?  Too bad it did not happen.  When Tony took the particle physics course, Persis was not teaching it.}

\indent{Now that I have mentioned Drell-Yan.  I better say something about it before my time is up.}

\section*{I. The Naive Model}

\indent{When I started thinking about what to say in this talk a little over a month ago, I discovered quickly that there was a vast amount of literature on the subject, and I did not know where to begin.  For help I contacted an old friend and one of the SLAC alumni Dave Soper.  As a good place to start, Dave recommended a 1995 review paper [1] and a 1996 book [2].  I am grateful to Dave for  for his excellent advice.  I also found two earlier reviews [3,4] which are very helpful.  For much of what I will say  below,  I have consulted these review papers.  I would also like to apologize in advance for not mentioning all the names of those making significant contributions to the field.  There are just too many.}

\indent{The field on lepton pair production began with the experiment at BNL by Christenson, Hicks, Lederman, Limon, Pope and Zavatini [5].  They studied the reaction
\begin{equation}
p + U \to \mu^{+}\mu^{-} + X
\end{equation}
for proton energies 22-29GeV, and the muon pair mass 1-6.7GeV.  The
results are shown in Fig. 1. 
  Two features of the data
 stand out: (1) the shoulder-like structure near the muon pair mass of 3GeV, and (2) the rapid fall-off of the cross section with the muon pair mass.  We now know that the shoulder-like structure is due to the $J/\psi$ which was discovered in 1974 by a muon pair production experiment at BNL [6] and an $e^{+}e^{-}$ colliding beam experiment at SLAC [7].The lepton pair production experiment also led to the discovery of the $\Upsilon$ family resonances in 1977 at Fermilab [8].  However, I would concentrate my talk on the continuum.}

\indent{Sid and I got interested in the process (1) for two reasons: (1) we were looking for application of the parton model outside deep inelastic lepton scatterings, and (2) we wanted to understand if the rapid decrease of the cross section with the muon pair mass could be reconciled with the point-like cross sections observed in the deep inelastic electron scattering.}

\indent{The key idea in our approach was the impulse approximation.  First, we picked an appropriate infinite momentum frame to exploit the time dilation.  In this frame, if we were able to establish that the time duration for the external current probe  $\tau_{{\rm probe}}$ is much shorter than the lifetimes of the relevant intermediate states $\tau_{{\rm int.states}}$, i.e.   
\begin{equation}
\tau_{{\rm probe}} << \tau_{{\rm int.states}}
\end{equation}
then the constituents could be treated as free.  Thus, the cross section in the impulse approximation is a product of the probability to find the particular parton configuration and the cross section for the free parton(s).  In the case of lepton pair production from two initial hadrons
\begin{equation}
P_{1}+P_{2}\to \ell^{+}\ell^{-} + X
\end{equation}
the pair production by the parton-antiparton annihilation satisfies the criteria of impulse approximation.  Another mechanism of the pair production by bremsstrahlung clearly does not.  From Fig. 2 it is easily shown that the fractional longitudinal momenta of the annihilating partons satisfy
\begin{equation}
\tau = x_{1}x_{2} = \frac{Q^{2}}{s}
\end{equation}
where $Q^{2}$ and s are respectively the pair mass squared and the square of the C.M. energy of the initial hadrons.  The rapidity of the pair is given by
\begin{equation}
y = \frac{1}{2}ln\frac{x_{1}}{x_{2}}
\end{equation}

The predictions stated in our original paper [9] are\\
\indent{1. The cross section $Q^{4}\frac{d\sigma}{dQ^{2}}$ depends only on the scaling variable $\tau = Q^{2}/s$;\\}
\indent{2. The magnitude and shape of the cross section are determined by the parton and antiparton distributions measured in deep inelastic lepton scatterings;}
\begin{equation}
\frac{d\sigma}{dQ^{2}dy} = \frac{4\pi \alpha^{2}}{3Q^{4}}(\frac{1}{N_{c}})\Sigma_{P}x_{1}f_{P}(x_{1})x_{2}f_{\bar{P}}(x_{2})
\end{equation}

where I have included a color factor $N_{c}$ in anticipating QCD.\\
\indent{3. If a photon, pion, kaon, or antiproton is used as the projectile, its structure functions can be measured by lepton pair production.  This is the only way I know of\\ to study the parton structure of a particle unavailable as a target;\\}
\indent{4. The transverse momentum of the pair should be small ($\sim$300-500 MeV);\\}
\indent{5. In the rest system of the lepton pair, the angular distribution is l + cos$^{2}\theta$ with respect to the hadronic collision axis, typical of the spin $\frac{1}{2}$ pair production from a transversely polarized virtual photon;\\}
\indent{6. The same model can be easily modified to account for W$^{\pm}$ boson productions.}\\

In this model, the rapid decrease of the cross section with Q$^{2}$ as seen in [1] is related to the rapid fall-off of structure functions as $x\to 1$ in the deep inelastic electron scatterings.}

\indent{The lepton-pair production that Sid and I considered was the first example of a class of hard processes involving two initial hadrons.  These processes are not dominated by short distances or light cone.  So the standard analysis using operator product expansion is not applicable.  But the parton model works.  Soon after our work, Berman, Bjorken and Kogut [10] applied similar ideas to large transverse momentum processes:
\begin{equation}
h_{1}+h_{2}\to h(large P_{T})+ X
\end{equation}
induced by deep inelastic electromagnetic interactions.  At that time, it was believed that strong interactions severely suppressed large transverse momenta, therefore electromagnetic interactions would quickly dominate the large transverse momentum processes.  This was the precursor of the point-like gluon exchanges in QCD.}

\section*{II. The Successor- QCD}

\indent{After the advent of QCD, the basic physical picture of the naive Drell-Yan model has been confirmed theoretically and the details have been greatly improved.  It is no longer a model.  That lepton pairs are produced by parton-antiparton annihilation is a consequence of QCD.  In QCD, the partons are quarks, antiquarks and gluons, and the number of color is three $N_{c}=3$.  The unique property of QCD being an asymptotically free gauge field theory makes the parton model almost correct, namely for deep inelastic processes we have
\begin{equation}
QCD = parton\, model + ``small''\, corrections
\end{equation}

\indent{The quotation marks on the right hand side are to indicate in one aspect to be mentioned below that the correction is unusually large.  In the modern language, the impulse approximation is replace by the more precise concept of factorization which separates the long distance and short distance physics and the condition (2) now becomes
\begin{equation}
Q^{2} >> \Lambda^{2}_{QCD}
\end{equation} 
The constituents are almost free leading to logarithmic corrections to the structure functions}
\begin{equation}
f_{i}(x)\Longrightarrow f_{i}(x,lnQ^{2}).
\end{equation}

\indent{The theoretical machinery to deal with the deep inelastic lepton scatterings is now familiar to everyone.  Predictions are obtained by combining QCD with modern technology in quantum field theory--renormalization group approach and operator product expansions [11].  The relation of this formal approach to the more intuitive parton model has become clear after Altarelli and Parisi and Gribov and Lipatov[12] elucidated the qualitative picture proposed by Kogut and Susskind[13]. These authors showed that the renormalization group equations and operator product expansions are equivalent to a set of evolution equations in $Q^{2}$ for $Q^{2}$ dependent quark and gluon distributions.  For the case of lepton pair production factorization works in a more complicated manner and it has taken the hard work of many people and several years to establish.  The main complication arises from the new feature of initial and final state interactions between the hadrons.  I will refer the interested reader to the reviews[13].  The result is fairly simple to state}
\begin{equation}
\frac{d\sigma^{AB}}{dQ^{2}dy} = \Sigma_{A,B}\int^{1}_{x_{A}}d\xi_{A}\int^{1}_{x_{B}}d\xi_{A}f_{a/A}(\xi_{A},Q^{2})f_{b/B}(\xi_{B},Q^{2})
H_{ab},
\end{equation}
where the sum over a and b are over parton species.  The parton distribution functions $f$ are the same as in deep inelastic lepton scatterings.  The function $H_{ab}$ is the parton level hard scattering cross section computable in perturbative QCD and is often written as 
\begin{equation}
H_{ab} = \frac{d\hat{\sigma}}{dQ^{2}dy}.
\end{equation}
 
To order $O(\alpha_{s})$ the QCD corrections are shown in Fig. 3.  Diagrams Fig 3 (b) and (c) can give rise to large transverse momentum lepton paris.  Explicit calculations have been done to two loops[14] and the factorization has been verified to this order.  But the proof of factorization has been carried out to all orders in $\alpha_{s}$[13].

\indent{The QCD improved predictions can be summarized as follows:\\
\indent{1. The logarithmic corrections in Q$^{2}$ can be absorbed by Q$^{2}$-dependent quark and antiquark distribution functions of the hadrons that appear in deep inelastic lepton scatterings with the rule of substitution Q$^{2}\to |Q^{2}|$.  Scaling is violated, but only logarithmically and in a way calculable.}\\
\indent{2. Analytic continuation from space-like q$^{2}$ (deep inelastic scattering) to time-like q$^{2}$ (lepton pair production) and the difference in kinematics between the two processes produce a nonleading finite correction with a very large coefficient.  The result is simplest in terms of moments[15]
\begin{equation}
\sigma_{n} = \sigma_{n}^{(0)}[1 + \frac{\alpha_{s}}{2\pi}\cdot\frac{4}{3}\pi^{2}+ ...]
\end{equation}
The $\pi^{2}$ term is unusually large.  For $\alpha_{s}\sim$ 0.2-0.3, the correction factor is about two.  This is the K factor.  The $\pi^{2}$ terms exponentiate[16]
\begin{equation}
1 + \frac{\alpha_{s}}{2\pi} \frac{4}{3} \pi^{2}\to exp (\frac{\alpha_{s}}{2\pi} \frac{4}{3} \pi^{2})
\end{equation}
This is known as the ``leading $\pi$ summation.''\\
\indent{3. A large transverse momentum of the lepton pair can be produced by recoil of quarks or gluons as shown in Fig. 3.  A simple dimensional analysis gives}
\begin{equation}
<k_{T}^{2}> = a + \alpha_{s}(Q^{2})sf(\tau,\alpha_{s})
\end{equation}
The constant a is related to the primordial or intrinsic transverse momenta of the quarks and antiquarks.\\
\indent{4. The full angular distributions in both $\theta$ and $\phi$ depend on input quark and gluon densities and are rather complicated[17].  For small $k_{T}$ the $\theta$ dependence is close to 1 + cos$^{2}\theta$ even when high order corrections are taken into account.  For large $k_{T}$, the $\theta$ dependence is expected to be substantially modified[18].}

\section*{III. Comparison between Theory and Experiments}

\indent{I will concentrate on four areas: scaling, absolute rate, transverse momentum, and angular distribution.}

{1. Scaling}\\

\indent{The simplest test of the theory is the scaling prediction that $Q^{d4}\frac{d\sigma}{dQ^{2}dy}$ depends only on the variable $\tau = Q^{2}/s$.  This is born out very well from both proton and pion data as shown in Figs. 4,5,6,[19,20,21].  These data show that the logarithmic violation of scaling is not very significant.}

{2. Absolute Rates}\\

\indent{Fig. 7[22] shows the Fermilab data from E605 Collaboration for 800GeV protons on a Cu target, and the theoretical prediction including the next-to-leading order corrections using the MRS(A) parton distribution[23].  Fig. 8[24] shows the comparison between the prediction and the CDF data at $\sqrt{s}$ = 1.8TeV.  The Z boson peak is prominatly visible.  The $O(\alpha^{2}_{s})$ theoretical predictions for the W and Z cross sections are compared with measurements from $p\bar{p}$ colliders in Fig. 9[25].  The agreement between theory and experiment is very good.  This is a nontrivial test of the $Q^{2}$ evolution of the parton distributions, since the input structure functions have Q$^{2}$ typically about 10-10$^{2}$GeV$^{2}$, but the collider data have Q$^{2}$ around 10$^{4}$GeV$^{2}$.}

{3. Transverse Momentum}\\

\indent{The data for the transverse momenta of the lepton pairs are shown in
 Figs. 10[26] and 11[27].  There is a clear increase in the transverse momentum with s.}
 
{4. Angular Distributions}\\

\indent{For the angular distribution we will only show the comparison in Fig. 12 between data for the Z decays from the CDF collaboration and the theory[28].  The agreement is good.}

\section*{IV.  Lepton Pair Production as a Physics Tool}

\indent{The process of lepton pair production is so well understood in perturbative QCD that it has now become an important and powerful tool in search of new physics information.  Let me mention a few below.\\
\indent{1. It played a crucial role in the design of the experiments at CERN which discovered the W and Z bosons.}\\
\indent{2. For the first time we have gotten a glimpse for how the quarks and antiquarks are distributed inside the unstable particles pion and kaon[29].}\\
\indent{3. By combining proton and antiproton data, it is possible to separate the valence and sea distributions inside a proton and an antiproton[30].}\\
\indent{4. Lepton pair production data have become an integral component of the global fits together with the deep inelastic lepton scatterings in determining the parton distributions inside a nucleon[1].}\\
\indent{5. For lepton pair production above the Z mass the electroweak interference is important.  The forward-backward-asymmetries can be utilized to be a efficient diagnostics tool to find Z$^\prime$ bosons if they exist[31].}\\
\indent{6. Lepton pair production with a polarized target/beam has been suggested[32] as a tool to pin down the spin-dependent parton distributions in a nucleon, especially for the sea quarks}

\indent{To illustrate the power of the lepton pair production as a physics tool, I will offer two more examples, the precision measurements of the W boson mass, and the lepton and quark sizes.}

\indent{Consider the W mass first.  The data quoted below come from the plenary talk by Dean Karlen[33] just a few days ago at the International High Energy Physics Conference in Vancouver.  The most precise measurements of the W mass at the Tevatron using lepton pair production and LEP2 are  summarized in Fig. 13, which also includes results from indirect measurements.  The W mass, top mass and Higgs mass in the Standard Model are all related through radiative corrections.  The interrelations are exhibited in Fig. 14.  We see that from our present knowledge on the W and top mass, the Higgs mass is interestingly constrained. The constraint will be tightened even more when the precision of the W mass improves.  In the future when the Tevatron completes Run 2 the uncertainty of the W mass will reduce to order of 40MeV.  It also has a program called TeV33 to increase the luminosity to 10$^{33}$.  When this happens, the uncertainty of the W mass is expected to shrink further to order of 20MeV.}

\indent{The second example concerns CDF experiment[34] on the sizes of leptons and quarks from the lepton pair production.  Using a model by Eichten, Land and Peskin with the Lagrangian[35]
\begin{equation}
L_{\pm} = \pm \frac{4\pi}{(\Lambda_{LL}^{\pm})}^{2} (\bar{E}_{L}\gamma^{\mu}E_{L})(\bar{Q}_{L}\gamma_{\mu}Q_{L}),~~
E = (\nu e,\nu_{e}),~~
Q = (u,d)
\end{equation}
CDF is able to show the composite scale parameters $\Lambda_{LL}^{\pm}$ to be in the range of 2.5 to 4.2 TeV.  The experimental data, the standard model predictions and the effects due to the new interactions(15) are shown in Fig.15.  The data obviously are consistent with the standard model predictions.}

\section*{V. Conclusions}

\indent{Since the first experiment at BNL and the naive model proposed to understand it, both experiments and theory have come a long way.  It is interesting to note that our original crude fit[9] did not even remotely resemble the data.  Sid and I went ahead to publish our paper because of the model's simplicity and our belief that future experiments would be able to definitively confirm or demolish the model.  It is gratifying to see that the successor of the naive model, the QCD improved version, has been confirmed by the experiments carried out in the last 28 years.  Lepton pair production process has been an important and active theoretical arena to understand various theoretical issues such as the infrared divergences, and collinear divergences leading to the factorization theorem in QCD for hard processes involving two initial hadrons.  The process has been so well understood theoretically that it has become a powerful tool for precision measurements and new physics.  We can expect to find new applications for using this process.}    

\section*{Acknowledgment}

\indent{This work is supported in part by the National Science Foundation}

\pagebreak

\newpage

\begin{flushleft}
 {\bfseries List of Figures}
\end{flushleft}

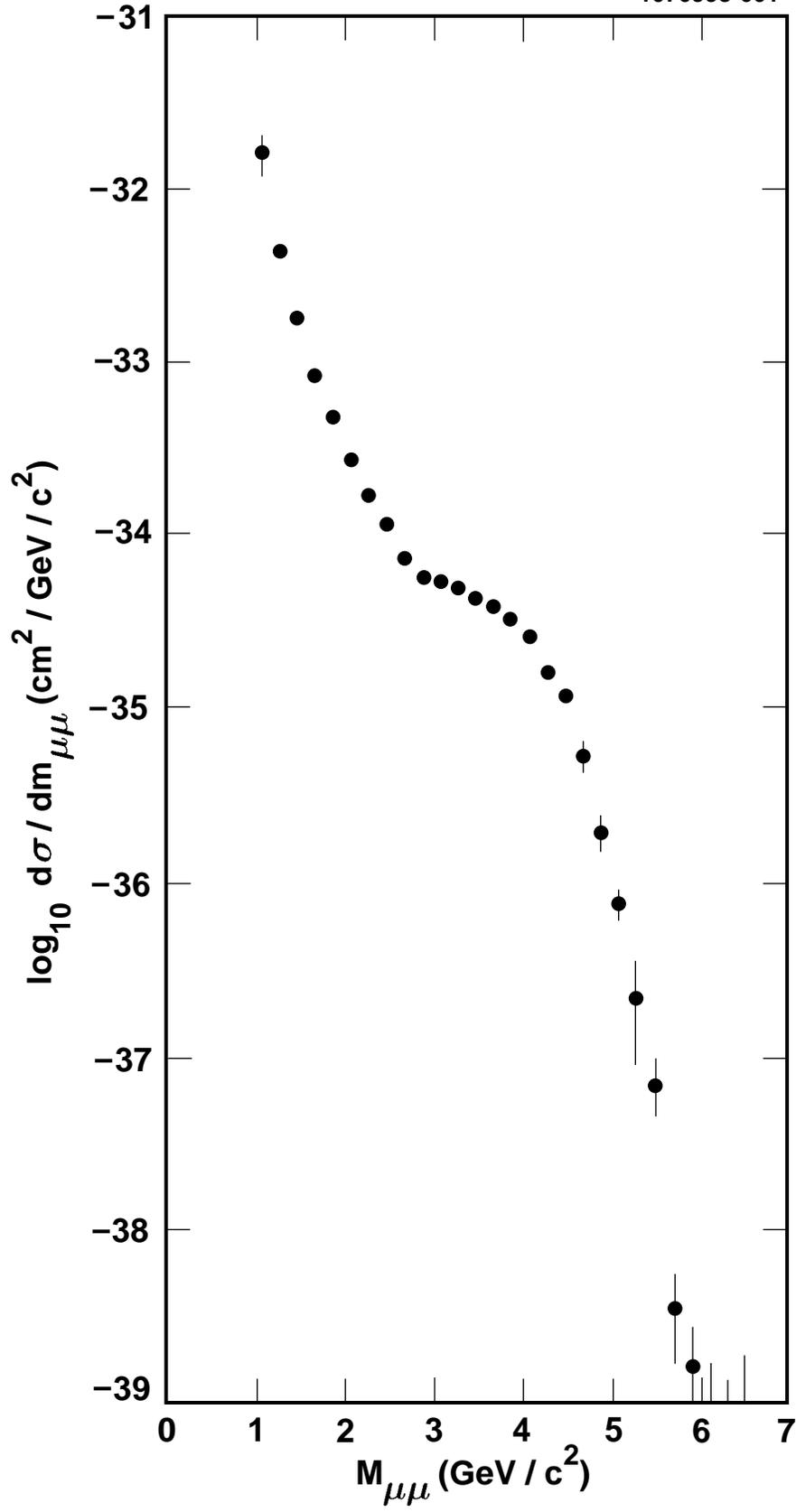
{\bf Fig. 1}$\;\;\;\;$
\begin{minipage}[t]{.7\textwidth}
   The dimuon spectrum from the BNL experiment
\end{minipage}

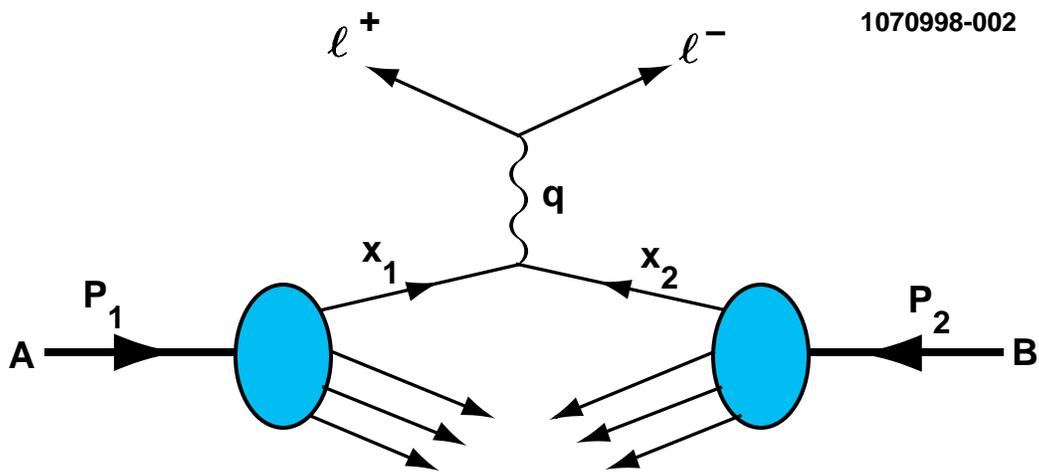
{\bf Fig. 2}$\;\;\;\;$\begin{minipage}[t]{.7\textwidth}The parton-antiparton annihilation mechanism for lepton pair production
\end{minipage}

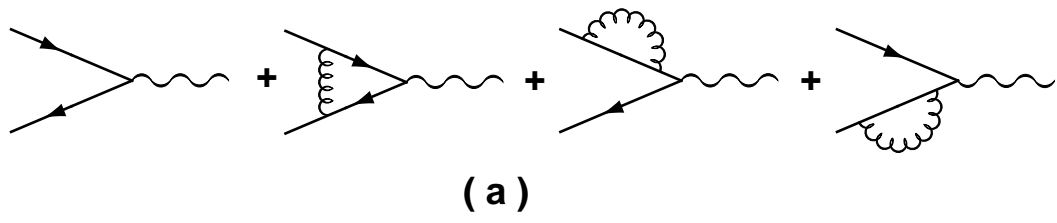
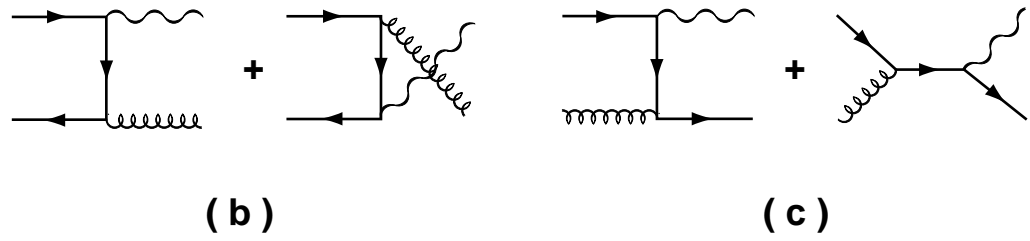
{\bf Fig. 3}$\;\;\;\;$\begin{minipage}[t]{.7\textwidth}Diagrams contributing to the $O(\alpha_{s})$ corrections for lepton pair production
\end{minipage}

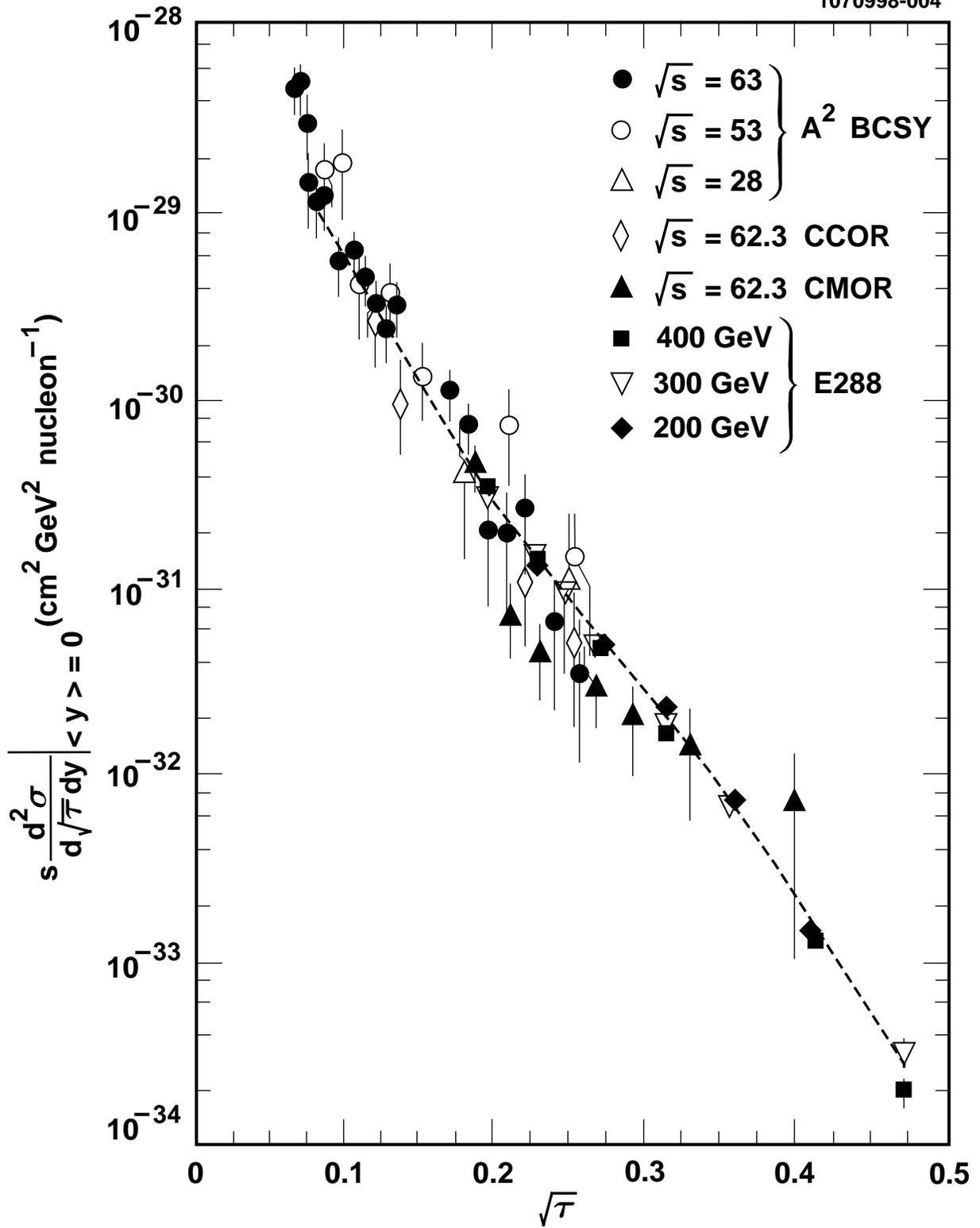
{\bf Fig. 4}$\;\;\;\;$\begin{minipage}[t]{.7\textwidth}Scaling of proton induced dilepton production (Ref.[19]). The curve is the prediction(6) scaled up by a factor of order 3.
\end{minipage}

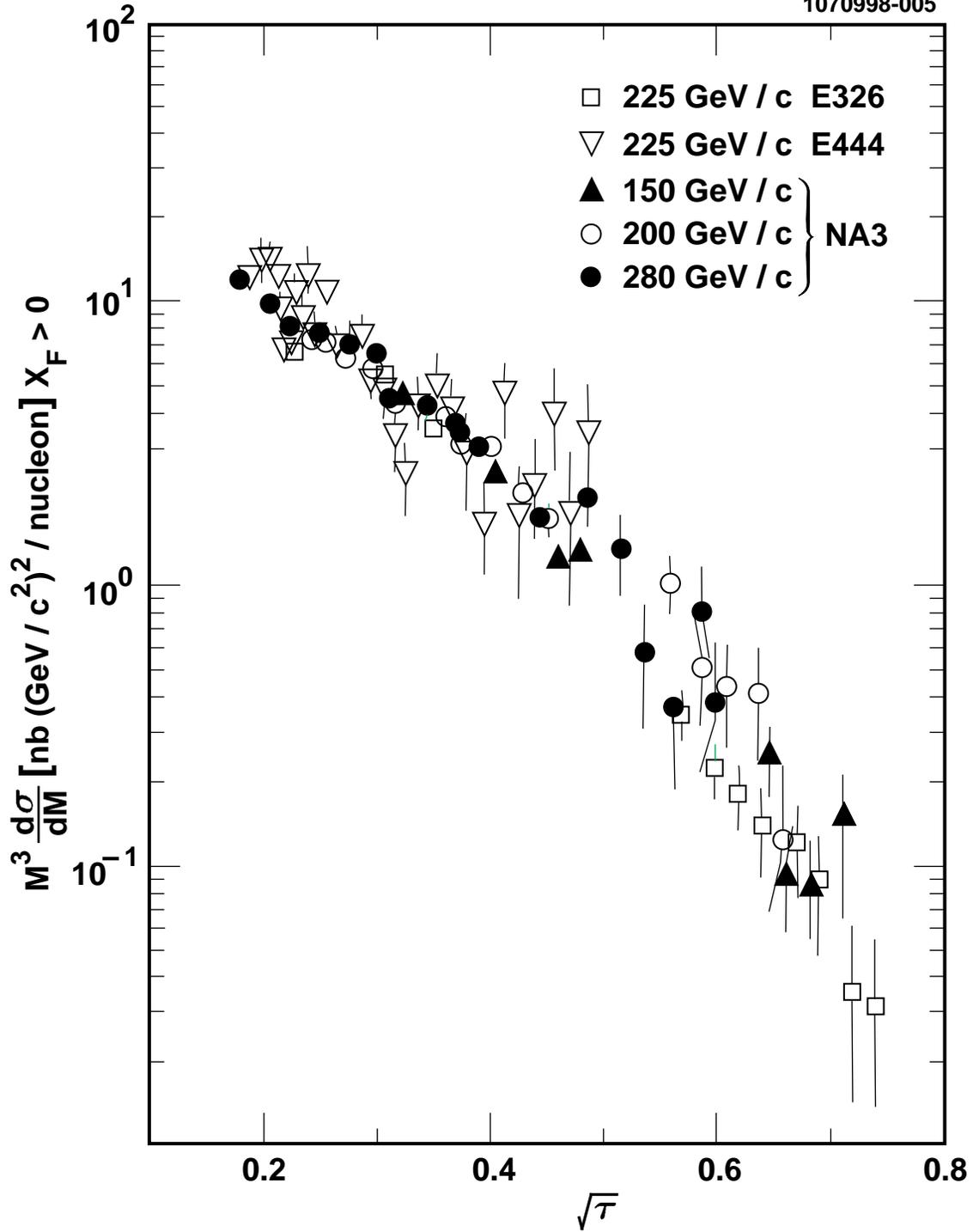
{\bf Fig. 5}$\;\;\;\;$\begin{minipage}[t]{.7\textwidth}Scaling of pion induced dimuon cross section(Ref.[20])
\end{minipage}

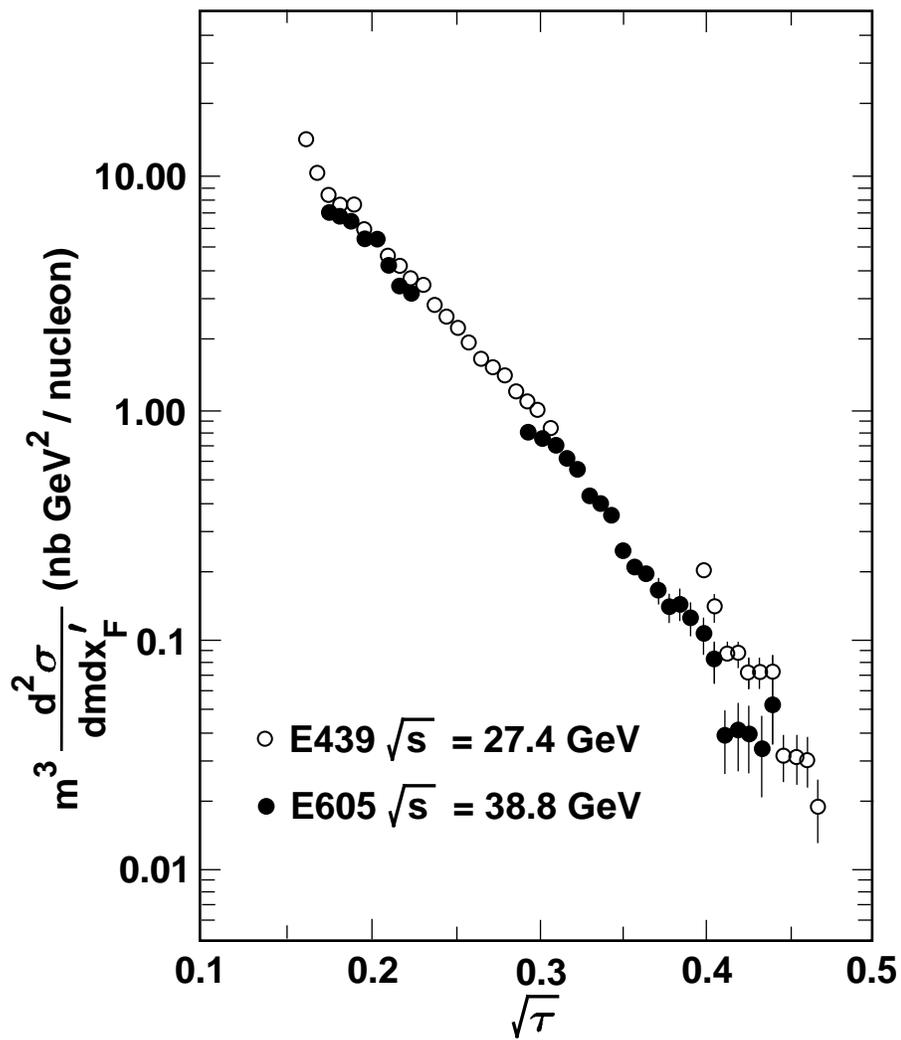
{\bf Fig. 6}$\;\;\;\;$\begin{minipage}[t]{.7\textwidth}Scaling form of the dimuon yield comparing data from E605 and E439 for the interval $0 < x^\prime_{F} < 0.2$ (Ref.[21])
\end{minipage}

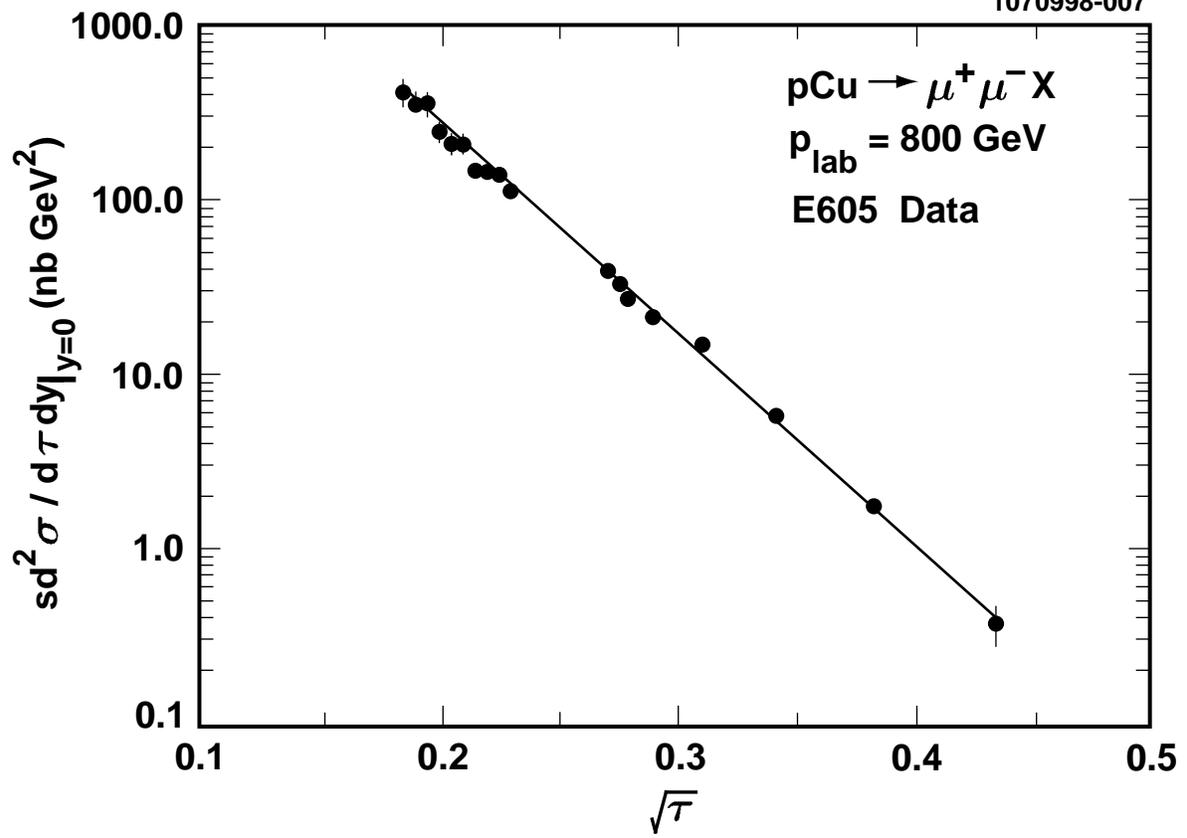
{\bf Fig. 7}$\;\;\;\;$\begin{minipage}[t]{.7\textwidth}The lepton pair production cross section measured by the E605 Collaboration(Ref.[21]) compared with the next to leading order theoretical prediction(Ref.[23])
\end{minipage}

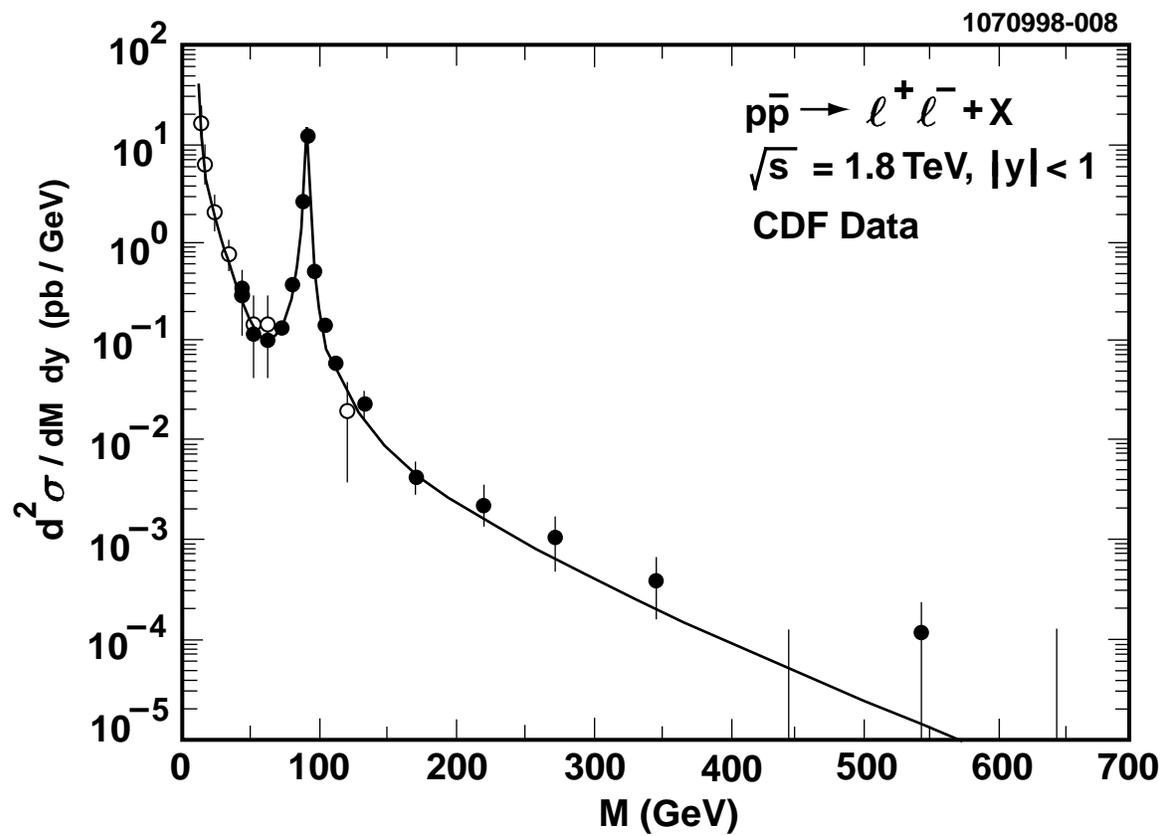
{\bf Fig. 8}$\;\;\;\;$\begin{minipage}[t]{.7\textwidth}The lepton pair cross section in p$\bar{p}$ collisions, with CDF data from Ref.[24(a)] (open circles) and Ref.[24(b)](solid circles).  The curve is the next to leading order QCD predictions using the parton distributions from Ref.[23]
\end{minipage}

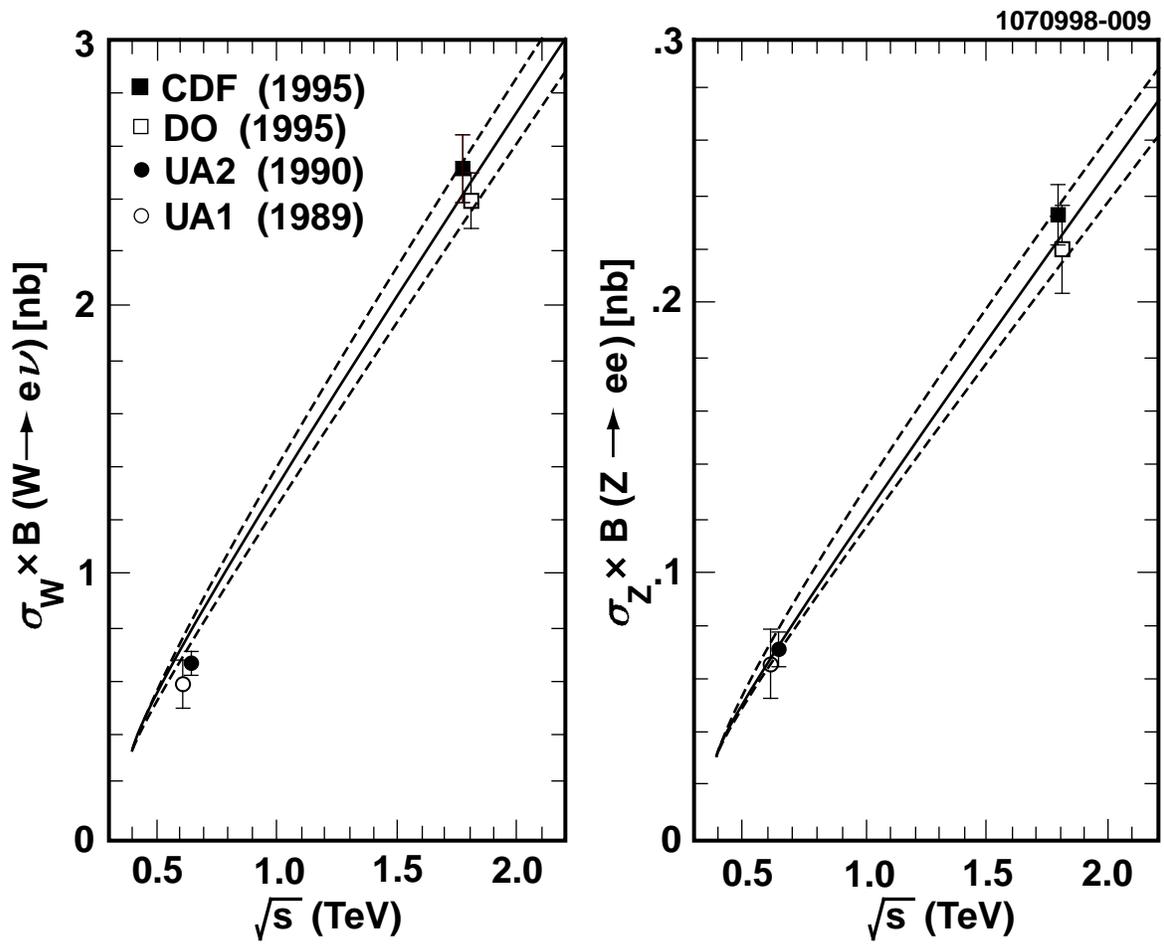
{\bf Fig. 9}$\;\;\;\;$\begin{minipage}[t]{.7\textwidth}Comparison of W and Z cross section measurements in p$\bar{p}$ collisions with theoretical predictions
\end{minipage}

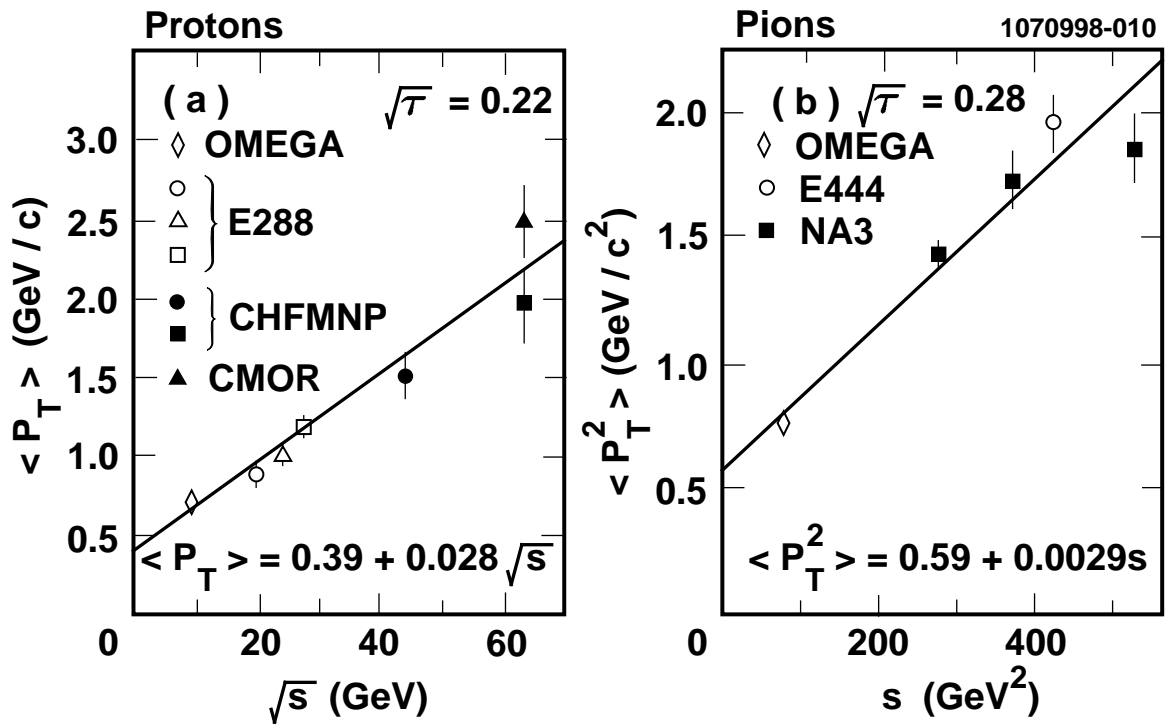
{\bf Fig. 10}$\;\;\;\;$\begin{minipage}[t]{.7\textwidth}(a) The mean transverse momentum for proton-induced lepton pairs.  (b) The mean squared transverse momentum for muon pairs produced in pion collisions
\end{minipage}

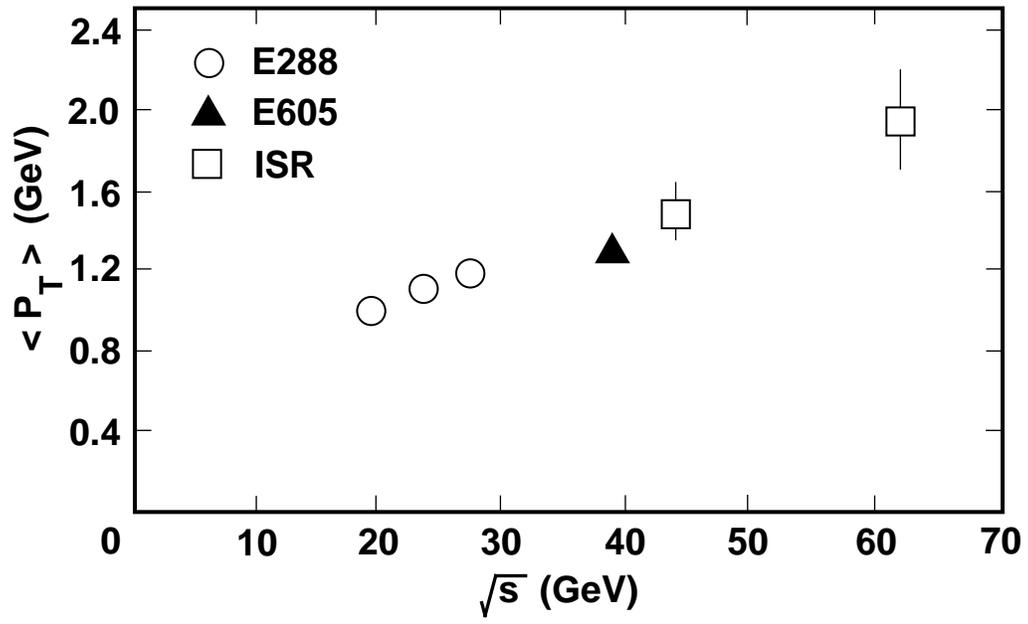
{\bf Fig. 11}$\;\;\;\;$\begin{minipage}[t]{.7\textwidth}The mean transverse momentum of muon pairs at $\sqrt{\tau} \simeq 0.3$ versus $\sqrt{s}$
\end{minipage}

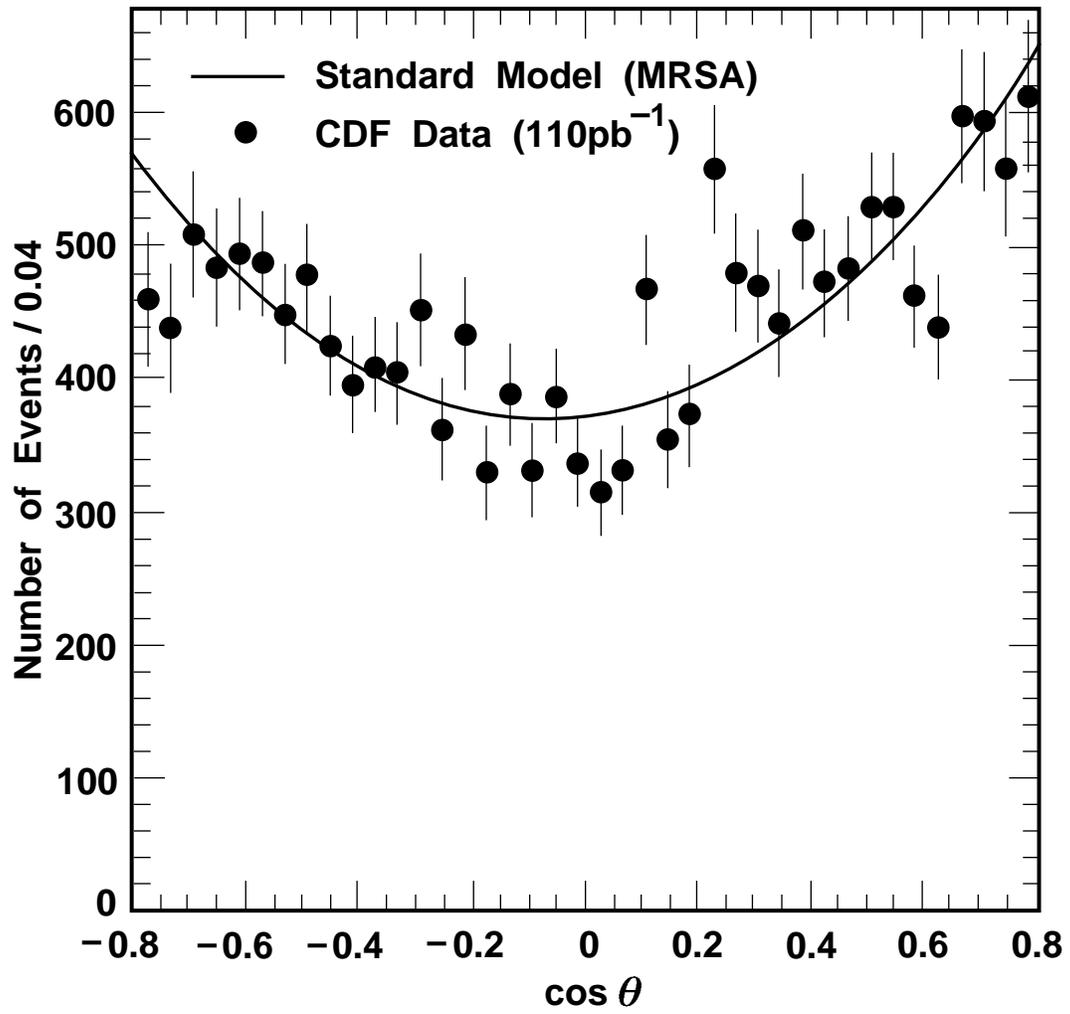
{\bf Fig. 12}$\;\;\;\;$\begin{minipage}[t]{.7\textwidth}Angular distribution of leptons from Z boson decay measured by CDF Collaboration
\end{minipage}

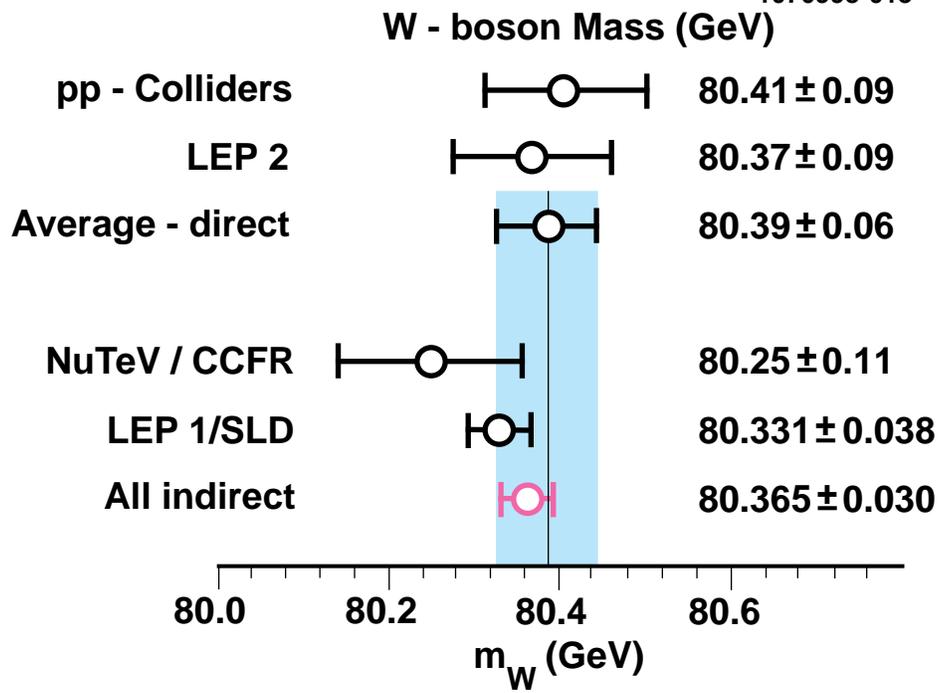
{\bf Fig. 13}$\;\;\;\;$\begin{minipage}[t]{.7\textwidth}Summary of Recent W boson mass measurements.  Figure is from Ref.[33]
\end{minipage}

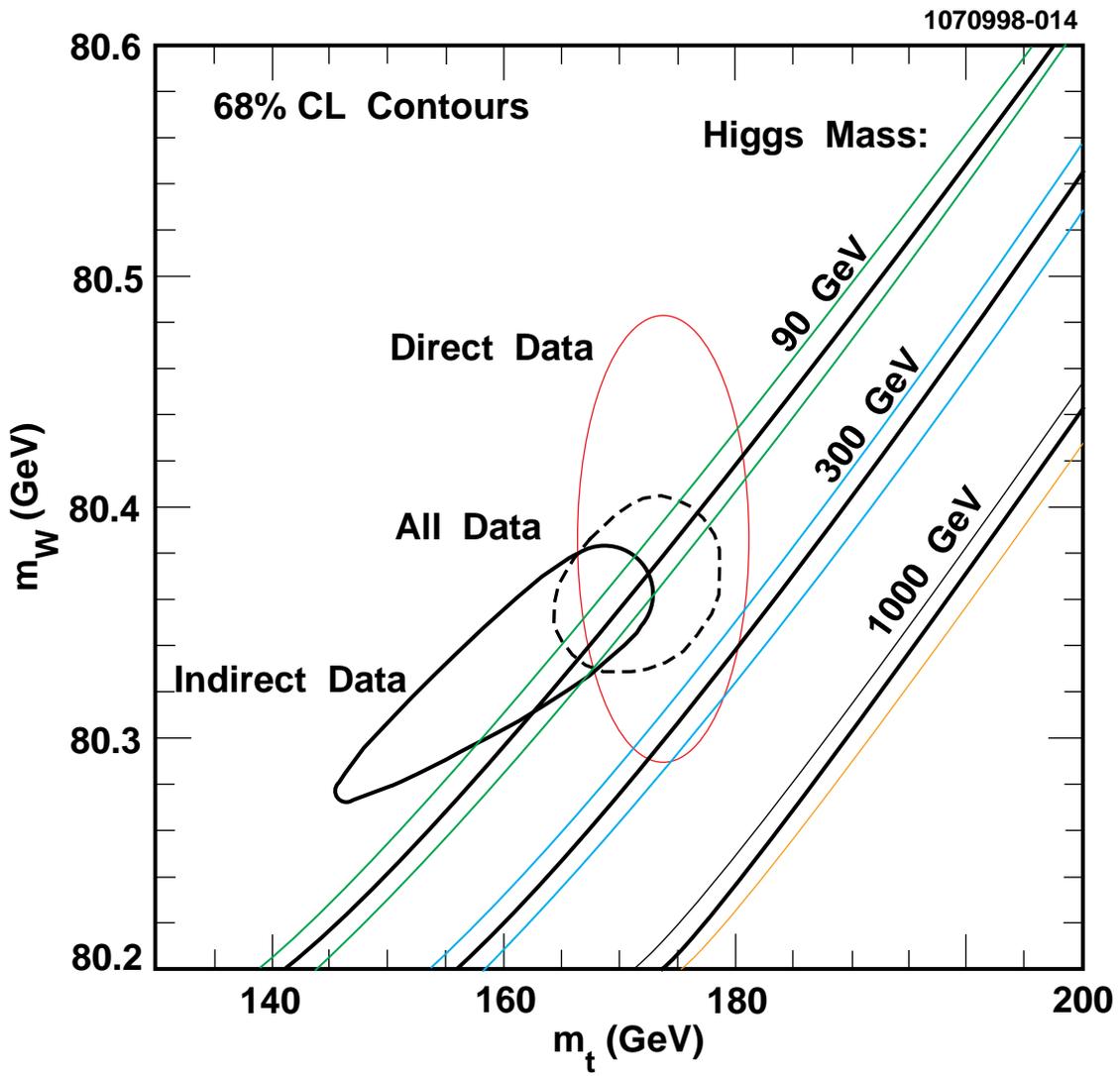
{\bf Fig. 14}$\;\;\;\;$\begin{minipage}[t]{.7\textwidth}Interrelations among W boson mass, top quark mass, and Higgs boson mass from recent data. Figure is from Ref.[33]
\end{minipage}

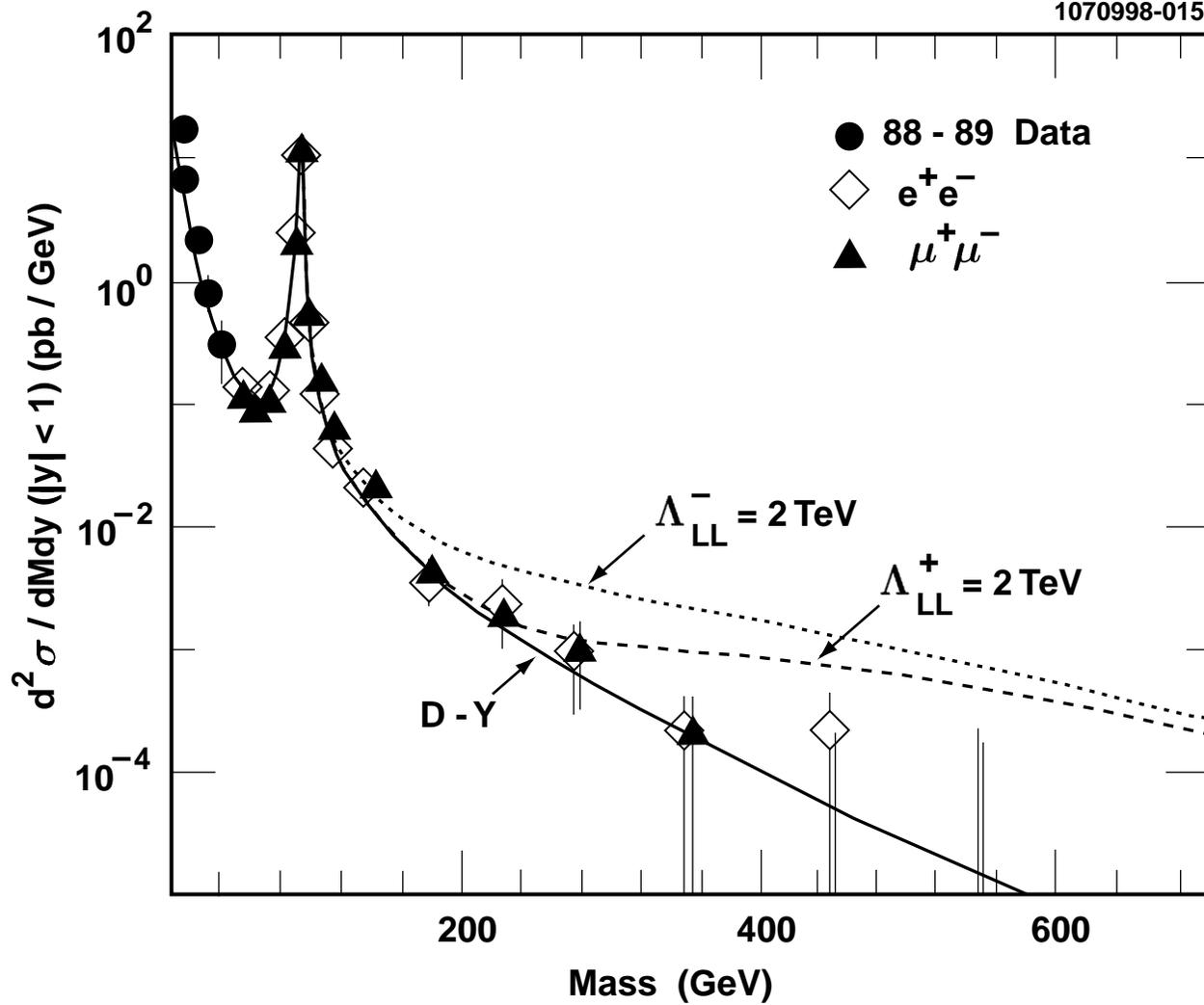
{\bf Fig. 15}$\;\;\;\;$\begin{minipage}[t]{.7\textwidth}Lepton pair production with $|y| < 1$ for $p\bar{p}\to \ell^{+}\ell^{-} + X$. The circles (M $<$ 50GeV) are from earlier data[37]; the data from 50 to 150GeV are normalized to the standard model value.  The curve is a standard model calculations based on the Lagrangian (15) with $\Lambda_{LL}$=2TeV
\end{minipage}

\newpage

\end{document}